\documentclass[aps, reprint,prl,nofootinbib,longbibliography]{revtex4-1}
\usepackage{dcolumn}
\usepackage[utf8]{inputenc}
\usepackage{amsmath}
\usepackage{amsfonts}
\usepackage{amssymb}
\usepackage{graphicx}
\usepackage{hyperref}
\usepackage{xcolor}

\newcommand{\ket}[1]{{\left|{#1}\right\rangle}}
\newcommand{\bra}[1]{\ensuremath{\left\langle{#1}\right |}}

\newcommand{\mb}{\mathbf}

\begin{document}


\title{Quantum-Relativistic Anomalous Proper Time}




\author{Pablo L. Saldanha}\email{saldanha@fisica.ufmg.br} \affiliation{Departamento de F\'isica, Universidade Federal de Minas Gerais, Belo Horizonte, MG 31270-901, Brazil}
\author{Nelson Yokomizo}\email{yokomizo@fisica.umfg.br}\affiliation{Departamento de F\'isica, Universidade Federal de Minas Gerais, Belo Horizonte, MG 31270-901, Brazil}

\date{\today}

\begin{abstract}
We investigate surprising consequences of the hypothesis that gravity and proper time obey the quantum superposition principle. We show that, for a massive body prepared in a superposition of two positions and post-selected in a suitable final state, the elapsed proper time in another region of space can  assume an anomalous effective value lying outside the range allowed by classical general relativity. Remarkably, the effective elapsed proper time can even become negative, corresponding to an effective evolution backward in time. We also propose an experiment to observe an anomalous effective proper time associated with the internal degrees of freedom of a quantum particle in an interferometer under Earth’s gravitational field. The proposed experiment is, in principle, feasible with current technology, and its realization could provide the first experimental evidence of an interplay between quantum mechanics and general relativity beyond the effects of the Newtonian potential in flat spacetime.
\end{abstract}


\maketitle

The flow of time exhibits markedly distinct features in different physical regimes. In Newtonian mechanics, time passes at the same rate for all observers and is unaffected by physical systems. In the relativistic domain, the flow of time is recognized as dependent on both the state of motion of the observer and the configuration of the gravitational field. Each localized system is then endowed with its own notion of proper time, measured by a clock carried along its worldline, and the amount of proper time elapsed for a system can be manipulated by changing its motion or the distribution of masses around its path. Understanding how this dependence of proper time on the trajectory of the system and the external gravitational field manifests itself in the evolution of quantum systems remains an open problem at the interface of quantum theory and gravitation.

As a quantum system can be spatially delocalized, a single proper time is, in general, insufficient to describe its evolution. For a system moving in a superposition of paths, different amounts of proper time elapse along each path for the same variation of time as measured by some clock in the laboratory. Among the effects expected to result from such a variation in the flow of proper time are a universal mechanism for decoherence of extended quantum systems \cite{Zych:2011hu,Pikovski:2013qwa} and specific phase shifts in quantum interferometry in external gravitational fields (see \cite{Mieling:2025iyd} and references therein). Moreover, a mass prepared in a superposition of positions may produce a nonclassical configuration of the gravitational field. A nearby quantum system would then experience a superposition of gravitational potentials and evolve under a corresponding superposition of proper times. Phenomenological models formulated to describe this regime predict possible low-energy manifestations of quantum gravity \cite{Bose:2024nhv,Marletto:2024ltk}, including gravitationally induced entanglement \cite{Bose:2017nin,Marletto:2017kzi}, processes with indefinite causal order \cite{Zych:2017tau,Moller:2023iwo}, and a repulsive behavior of the gravitational force \cite{Saldanha:2026zar,Bose:2026tuv}, whose observation could provide evidence for a nonclassical nature of the gravitational field.

Here we discuss how the flow of time can behave in an unexpected way under the assumption that gravity obeys the quantum superposition principle. We consider the elapsed proper time in a region of space under the influence of a mass $M$ prepared in a superposition of two different positions relative to this region. According to general relativity, the proper time $\tau_A$ spent in the region when the mass $M$ is closer to this region is smaller than the proper time $\tau_B$ elapsed when the mass $M$ is farther away. We show that, with a suitable post-selection on the final state of the mass, the elapsed proper time in the region may assume an anomalous effective value lying outside the interval $[\tau_A, \tau_B]$ associated with classical mixtures of the proper times in the two branches in superposition, thus going beyond what can be accounted for by classical general relativity. Remarkably, the elapsed proper time can even become negative. This implies that manipulating a mass in one region of space can, with a finite probability, induce an effective backward evolution of a system in another region of space.

The same mechanism applies to the proper time associated with the internal degrees of freedom of a quantum particle in an interferometer under Earth’s gravitational field. We also propose an experiment that could observe an anomalous effective proper time in this setting. To date, no experiment has demonstrated a superposition of proper times. If realized, our proposal would constitute the first experimental observation of a genuinely general-relativistic quantum effect. Moreover, it appears to be feasible with current technology.

Let us first briefly review the dynamics of a quantum system in a curved spacetime with metric $g_{ab}$, which has been studied in several theoretical frameworks \cite{Kiefer:1990pt,Lammerzahl:1995zz,Giulini:2012zu,Schwartz:2018pnh,Zych:2011hu,Zych2017}. The leading corrections due to a nonvanishing gravitational field in the weak field limit can be described  
as follows \cite{Zych:2011hu,Zych2017}. Consider a quantum system containing internal and external degrees of freedom, represented by a quantum state $\ket{\Phi(t)}$, and time-evolution dictated by a Hamiltonian $H$ in a flat spacetime. For a nonvanishing Newtonian potential, the classical proper time is related to the coordinate time according to
\begin{equation}
\label{eq:time-dilation}
    \frac{d\tau}{dt} = \sqrt{-g_{00}} \simeq 1 + \frac{\phi}{c^2} \, ,
\end{equation}
where $\phi(\mathbf{x},t)$ is the Newtonian potential. At each region of space, the system evolves with respect to the local proper time, and the Schr\"odinger equation becomes
\begin{equation}
\label{eq:time-evolution}
    i \hbar \frac{\partial}{\partial \tau} \ket{\Phi(t)} = H \ket{\Phi(t)} \, .
\end{equation}
Using Eq. (\ref{eq:time-dilation}), we can express the evolution equation \eqref{eq:time-evolution} in terms of the coordinate time in the form
\begin{equation}
    i \hbar \frac{\partial}{\partial t} \ket{\Phi(t)} = H \left( 1 + \frac{\phi(\mathbf{x},t)}{c^2}\right)\ket{\Phi(t)} \, .
\end{equation}
Higher-order corrections in powers of $c^{-1}$ can also be considered, but here we restrict our attention to the leading-order corrections described above.

For a system of mass $m$, the Hamiltonian $H$ includes the relativistic rest-energy term $m c^2$. In the presence of a Newtonian potential, this term is multiplied by the time dilation factor given in Eq.~\eqref{eq:time-dilation}, so that $m c^2 \to m c^2+m\phi$. Hence, the gravitational potential term in the Schr\"odinger equation can be understood as arising from the effect of gravitational time dilation on the relativistic rest-energy term. The evolution of the internal degrees of freedom is likewise affected by gravitational time dilation. The influence of the gravitational potential on the Sch\"odinger dynamics has already been experimentally observed \cite{Colella:1975dq,Nesvizhevsky:2002ef}. This effect can, however, be interpreted simply as arising from the inclusion of the nonrelativistic gravitational potential term $m\phi$ in the Schr\"odinger equation in a flat spacetime. A superposition of proper times associated with the internal degrees of freedom, on the other hand, has not yet been experimentally demonstrated, and would constitute a genuinely general-relativistic effect in quantum mechanics that does not admit a simple interpretation within Newtonian gravity. In this work, we describe a new mechanism that provides a promising strategy for observing this effect.

Consider the concrete scheme illustrated in Fig.~1, consisting of a ``source'' massive quantum particle in an interferometer, with possible paths A and B, and a nearby ``probe'' quantum system. The two systems interact gravitationally. This scheme is similar to that used to obtain an effective repulsive gravitational force \cite{Saldanha:2026zar} due to a quantum interference of forces \cite{Correa:2018jhg,Cenni:2019vrk,Militani:2024jgp}. Here, however, we focus on the flow of time rather than on momentum exchanges. Let the quantum state of the source particle between times $t_1=0$ and $t_2$ in the laboratory frame be the superposition $\ket{\phi_i}=\alpha\ket{A}+\beta\ket{B}$ with $|\alpha|^2+|\beta|^2=1$, where $\ket{A}$  and $\ket{B}$ represent states localized around the trajectories corresponding to the paths A and B in Fig. 1. At time $t_2$ in the laboratory frame, the source particle is detected in one of the two exits of the interferometer.

\begin{figure}
    \centering
    \includegraphics[width=8.5cm]{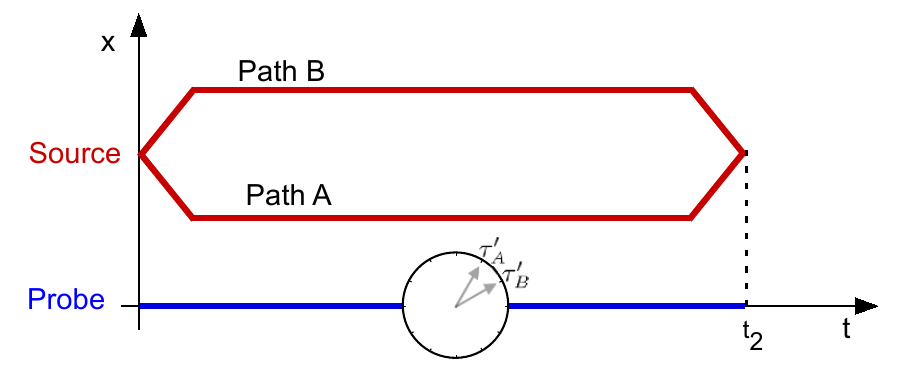}
    \caption{Scheme to obtain an anomalous proper time for the probe quantum system. 
    A  source massive  quantum particle has two possible paths in a 1D Mach-Zehnder interferometer. If the source mass propagates along path A, closer to the origin, a clock in the probe system runs slower than if the source mass propagates along path B, farther away. The clock pointer then becomes entangled with the position of the source mass in the superposition state depicted in the figure. A post-selection of the source mass at the exit of the interferometer can generate an effective elapsed proper time for the probe system outside what is permitted by general relativity. 
    }
 \end{figure}

Depending on the position of the source particle, the probe system is subjected to different gravitational potentials $\phi$. We assume that the source mass is sufficiently small and moves along the interferometer at  nonrelativistic speeds with $v/c \ll 1$, so that its gravitational field can be described by a Newtonian potential in the Newtonian limit of general relativity. Then, according to Eq. (\ref{eq:time-dilation}), if the source particle is in path A, the elapsed proper time $\tau_A$ of the probe system between times $t_1$ and $t_2$ in the laboratory frame is smaller than the elapsed proper time $\tau_B$ if the source particle is in path B.  

Let us assume that gravity can exist in a superposition state, such that the probe system evolves under a superposition of proper times. Under this assumption, if $\ket{\Psi_0}$ is the initial state of the probe system at time $t_1=0$, evolving through a Hamiltonian $H$, the composite state of the source particle and the probe system at time $t_2$ in the laboratory frame will be
\begin{equation}\label{state1}
\ket{\Psi_f}= e^{-\frac{i}{\hbar}H\tau_A}\left[\alpha\ket{A}+\beta\ket{B} e^{-\frac{i}{\hbar}H(\tau_B-\tau_A)}\right]\ket{\Psi_0}.
\end{equation}
where $\tau_A$ and $\tau_B$ are the proper times elapsed along the paths A and B, respectively, between the coordinate times $t_1$ and $t_2$. Note that in general there will be a gravitationally induced entanglement between the source particle and the probe system, since the time evolution of the probe system depends on the position of the source mass. We consider that the elapsed proper times $\tau_A$ and $\tau_B$ are close enough so that we can approximate
\begin{equation}\label{time_ev}
	e^{-\frac{i}{\hbar}H(\tau_B-\tau_A)}\ket{\Psi(0)}\simeq\left[1-\frac{i}{\hbar}H(\tau_B-\tau_A)\right]\ket{\Psi_0}.
\end{equation}

Suppose that the source particle is detected at a particular exit of the Mach-Zender interferometer, corresponding to a projection onto the state $\ket{\phi_f}=1/\sqrt{2}(-\ket{A}+\ket{B})$.  It can then be readily shown that, under this post-selection and with the approximation of Eq. (\ref{time_ev}),  the quantum state of the probe system at time $t_2$ in the laboratory frame reads
\begin{eqnarray}\nonumber\label{post-sel}
\ket{\Psi_{ps}}&\simeq&\frac{\beta-\alpha}{\sqrt{2}}e^{-\frac{i}{\hbar}H \tau_A}\left[1-\frac{i\beta H(\tau_{B}-\tau_A)}{(\beta-\alpha)\hbar}\right]\ket{\Psi_0}\\
&\simeq& \frac{\beta-\alpha}{\sqrt{2}}e^{-\frac{i}{\hbar}H \tau_{ef}}\ket{\Psi_0},
	\end{eqnarray}
where
\begin{equation}\label{tau_ef}
\tau_{ef}=\tau_A+\frac{\beta(\tau_B-\tau_A)}{\beta-\alpha}=\frac{\beta\tau_B-\alpha\tau_A}{\beta-\alpha}
\end{equation}
represents the effective elapsed proper time for the probe system under the post-selection of the source particle state. To arrive at Eq. (\ref{post-sel}), it was assumed that Eq. (\ref{time_ev}) is valid with $\tau_{ef}$ replacing $\tau_B$. 

From now on, we assume that $\alpha$ and $\beta$ are both real and positive. In this case, it can be seen in Fig. 2 that $\tau_{ef}$ is always anomalous.  For $\beta>1/\sqrt{2}$ it is always larger than $\tau_B$, while for $\beta<1/\sqrt{2}$ it is always smaller than $\tau_A$. Therefore, $\tau_{ef}$ cannot be obtained as a statistical mixture of $\tau_A$ and $\tau_B$, since any such mixture would lie between $\tau_A$ and  $\tau_B$.

\begin{figure}
    \centering
    \includegraphics[width=7.5cm]{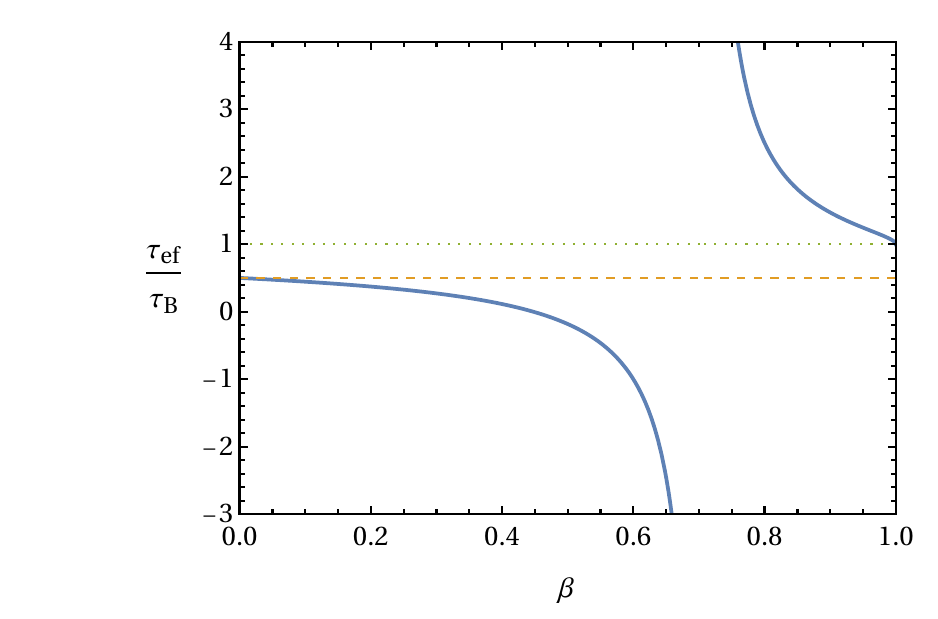}
    \caption{{Anomalous effective proper times.} The effective proper time $\tau_{ef}$ from Eq. (\ref{tau_ef}), normalized by $\tau_B$, is shown as a function of $\beta$, for $\tau_A=\tau_B/2$ and  $\alpha=\sqrt{1-\beta^2}$. $\tau_{ef}$ is always anomalous, exceeding $\tau_B$ (dotted green line) for $\beta>1/\sqrt{2}$ and falling below $\tau_A$ (dashed yellow line) for $\beta<1/\sqrt{2}$. It can even become negative, corresponding to an effective backward evolution in time.}
 \end{figure}

For $\beta<\alpha$ and $\beta\tau_B>\alpha\tau_A$, $\tau_{ef}$ from Eq. (\ref{tau_ef}) is negative, as shown in Fig.~2. The procedure depicted in Fig. 1 can be repeated many times. If in each procedure the source particle is successfully post-selected in the desired state, the probe quantum state will evolve through successive negative effective proper time intervals $\tau_{ef}$. In this case, the probe quantum system will effectively evolve backward in time. Any clock carried by the probe system would therefore run towards the past under these conditions.  One could wonder if the proposed scheme could act as a probabilistic time-reversal machine, reversing the temporal evolution of the probe quantum system with some probability of success. As an illustration, the probe system may consist of a closed box with a cat and a mortal device that kills the cat if an atom initially in an excited state decays to its fundamental state. Consider that at time $t=0$ in the laboratory frame the chance that the atom had decayed and the cat is dead is 50\%. If we succeed in performing the successive procedures with the correct post-selections, we could arrive at a situation where the atom has not decayed and the cat is alive with 100\% certainty. Would we resurrect Schrödinger's cat?

Although the resurrection of Schrödinger's cat would be a great achievement, there are problems with this interpretation. First, the evolution of the probe system with an anomalous proper time is probabilistic. The most probable post-selection of the source particle state at the exit of the interferometer of Fig. 1 is onto the state $1/\sqrt{2}(\ket{A}+\ket{B})$, which yields an effective elapsed proper time $(\beta\tau_B+\alpha\tau_A)/(\beta+\alpha)$, which lies between $\tau_A$ and $\tau_B$. Taking both possible results into account with their corresponding probabilities, the expected value of the effective elapsed proper time is the classical result, $\alpha^2 \tau_A + \beta^2\tau_B$. In the Schrödinger's cat example discussed above, the initial chance that the atom is in an excited state is $50\%$ and the probability of each successful post-selection is smaller than $50\%$. Therefore, there is no gain in the cat's life expectancy. A second limitation is that any correlation of the probe system and its surroundings, such as a dying meow sound escaping from the box, acts as a source of decoherence and prevents the system to evolve through an anomalous proper time. 
Thus, the presented scheme is not expected to work with macroscopic systems.

The effective backward time evolution discussed above should, however, work with microscopic isolated systems. Let the probe system be a quantum harmonic oscillator isolated from its environment, for instance. If the oscillator  mass is localized at the equilibrium position and moving to the right at time $t=0$ in the laboratory frame, if the desired post-selection is performed on the source particle in the scheme of Fig. 1, at the posterior time $t_2$ the expectation value of the mass position should be on the left side of the equilibrium position, moving to the right, as in a previous time. 

But there is a restriction for the probe system anomalous effective time evolution. If the time uncertainty of the probe system clock is $\Delta \tau$, there must be an energy uncertainty $\Delta E>\hbar/\Delta \tau$ in the probe system state. So, for Eq. (\ref{time_ev}) to be valid with $\tau_{ef}$ replacing $\tau_B$, an assumption that was used to obtain $\tau_{ef}$ from Eq. (\ref{tau_ef}), we must have $\Delta E |\tau_{ef}-\tau_A|/\hbar<1$, or $|\tau_{ef}-\tau_A|<\Delta\tau$. The difference between the anomalous effective time evolution and the usual one must be smaller than the clock time uncertainty. So, the effect cannot be observed with a single quantum system, an ensemble is necessary. The evolution of the probe system can be verified by measurements performed on identically prepared systems in the scheme of Fig. 1 at time $t_2$ in the laboratory frame. If the results of such measurements are consistent with the anomalous elapsed proper time from Eq. (\ref{tau_ef}), but not with an elapsed proper time between $\tau_A$ and  $\tau_B$, this behavior would provide evidence for a nonclassical nature of gravity, since it cannot be described as a classical mixture of the proper times described by general relativity. Instead, such a behavior is naturally explained by a quantum superposition of distinct states of gravity.

Note $\tau_{ef}$ from Eq. (\ref{tau_ef})  can be much greater than $\tau_B$  or much smaller than $\tau_A$ when $|\beta-\alpha|\ll1$, as shown in Fig. 2. This behavior is associated with the phenomenon of weak value amplification, and our results can be formulated within the weak value formalism \cite{Aharonov:1988xu,Dressel:2014kub}. The effective proper time $\tau_{ef}$ from Eq. (\ref{tau_ef}) can be written as the weak value of the operator $\hat{T}\equiv\tau_A\ket{A}\bra{A}+\tau_B\ket{B}\bra{B}$, associated with the elapsed proper time of the system for each interferometer arm of the source, for the pre-selected state $\ket{\phi_i}$ and the post-selected state $\ket{\phi_f}$: $\tau_{ef}=\bra{\phi_f}\hat{T}\ket{\phi_i}/\langle\phi_f|\phi_i\rangle$. The greater the weak value amplification, the lower the probability of post-selection, which is approximately given by $|\langle\phi_f|\phi_i\rangle|^2=|\beta-\alpha|^2/2$. 



It is very difficult to put massive particles in a superposition state at different locations. Even in the most optimistic scenarios for achieving such superpositions in the next decades \cite{Bose:2024nhv,Marletto:2024ltk}, the difference between the proper times $\tau_B$ and $\tau_A$ in the scheme of Fig. 1 remains completely negligible. As a result, although the scheme of Fig. 1 raises interesting questions regarding the possible quantum behavior of gravity and an associated anomalous flow of time, it appears less promising than other schemes proposed in the literature for witnessing a nonclassical nature of gravity \cite{Bose:2017nin,Marletto:2017kzi,Saldanha:2026zar,Bose:2026tuv,Bose:2024nhv,Marletto:2024ltk}. 

The scheme of Fig. 3, on the other hand, is more promising as a means of witnessing a quantum behavior associated with the gravitational field produced by a mass. Now, the ``source'' for the different proper times is Earth's gravitational field and the ``probe'' quantum system is the internal degree of freedom of a quantum particle in a superposition of different gravitational potentials in a Mach-Zehnder interferometer. Suppose that the spatial degree of freedom of the quantum particle is in the state $\alpha\ket{A}+\beta\ket{B}$ after the particle wave function is split by the beam splitter BS1, where $\ket{A}$ and $\ket{B}$ represent the interferometer paths $A$ and $B$ depicted in Fig. 3. We assume that the probe system is initially prepared in the state $\ket{\Psi_0}$ at time $t_1=0$ in the laboratory frame, just after the split by BS$_1$. If we assume, as before, that the probe system evolves in a superposition of different proper times $\tau_A$ and $\tau_B$ associated to each path, its final state at time $t_2$ in the laboratory frame is the same as represented in Eq. (\ref{state1}). If a post-selection on the state $1/\sqrt{2}(-\ket{A}+\ket{B})$ is performed, for instance with the particle leaving the interferometer by the exit E$_1$, then the quantum state of the probe system will be given by Eq. (\ref{post-sel}), with $\tau_{ef}$ given by Eq. (\ref{tau_ef}), under the same approximations. The same results previously discussed for the probe system in the scheme of Fig. 1 then apply to the internal degrees of freedom of the system in the scheme of Fig. 3.

\begin{figure}
    \centering
    \includegraphics[width=8.5cm]{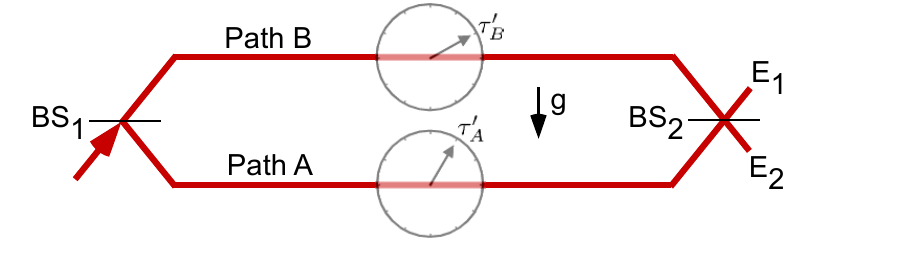}
    \caption{{Proposed scheme to verify an anomalous proper time.} A quantum particle with a internal degree of freedom working as a clock propagates in a Mach-Zehnder interferometer under Earth's gravitational field $\mb{g}$. BS$_1$ and BS$_2$ are beam splitters. Since the proper time for the particle internal degree of freedom runs faster for a higher gravitational potential, the clock pointer will be entangled with the particle path, in the superposition state depicted in the figure. Depending on the particular exit the particle leaves the interferometer, the elapsed proper time for the clock may be outside what is permitted by general relativity, what could be experimentally verified.}
 \end{figure}

For the scheme of Fig. 3, we have $(\tau_B-\tau_A)/\tau_B\approx gh/c^2$, where $g$ is Earth's gravitational field and $h$ is the average height difference between paths $A$ and $B$. For an interferometer with $h\approx 1$m, we obtain $(\tau_B-\tau_A)/\tau_B\approx10^{-16}$. We may have $(\tau_{ef}-\tau_B)/\tau_B$ greater than $10^{-16}$ for these parameters, if we use weak value amplification. Note that atomic clocks can have a precision of one part in $10^{19}$ \cite{Brewer:2019gyz,Marshall:2025ahl}. With such a precision, $\tau_{ef}$ can be distinguished from both $\tau_A$ and $\tau_B$.



The quantum system in the scheme of Fig. 3 can be a two-level atom, as in an atomic clock working with Ramsey interferometry \cite{Ludlow2015}. At time $t_1=0$, a light pulse from a laser with angular frequency $\omega$ induces a $\pi/2$ Rabi cycle in the atom initially in the ground state, preparing a state $\ket{\Psi_0}=(\ket{0}+\ket{1})/\sqrt{2}$, where $\ket{0}$ represents the atom ground state, with energy $0$, and $\ket{1}$ represents the excited state, with energy $\hbar\omega_0$. At time $t_2$, just after the paths recombination at BS$_2$ in Fig. 3, another light pulse from the same laser interacts with the atom, inducing another $\pi/2$ Rabi cycle. The probability that the atom ends in the excited state for an elapsed proper time $\tau$ for the atom and a elapsed proper time $\tau_l$ for the laser is $\cos^2(\theta/2)$, with $\theta=\omega_0\tau-\omega \tau_l$ \cite{grynberg}. Note that the atom clock accumulates a phase $\omega_0\tau$, while the laser field accumulates a phase $-\omega \tau_l$. So, the Ramsey interference phase is $\theta_{ef}=\omega_0\tau_{ef}-\omega \tau_l$ if the atom exits through the post-selection port, with an effective proper time $\tau_{ef}$. With the beam splitter BS$_2$ removed in the scheme of Fig. 3, the Ramsey interference phase is  $\theta_{A}=\omega_0\tau_{A}-\omega \tau_l$ if the atom is collected in the region E$_1$ and $\theta_{B}=\omega_0\tau_{B}-\omega \tau_l$ if the atom is collected in the region E$_2$. In this way, it could be verified whether $\tau_{ef}$ has an anomalous value, being greater than $\tau_B$ or smaller than $\tau_A$, through $\theta_{ef}-\theta_A=\omega_0(\tau_{ef}-\tau_A)$ and  $\theta_{ef}-\theta_B=\omega_0(\tau_{ef}-\tau_B)$. This would characterize a quantum superposition of different proper times in the evolution of the system, a behavior that cannot be explained within classical gravity. Considering, for instance, an optical transition with  $\omega_0/(2\pi)\approx 6\times10^{14}$Hz, an average height difference between paths $A$ and $B$ $h\approx 10$cm, an interval between the Ramsey pulses $t_2\approx1$s, and  $\beta=1/\sqrt{2}+0.1$ in Eq. (\ref{tau_ef}), we obtain $\theta_{ef}-\theta_B\approx0.1$, which in principle could be experimentally verified. 
In Ref. \cite{Kovachy:2015xcp} an atomic Mach-Zehnder interferometer with more challenging parameters ($h\approx 50$cm and $t_2\approx2$s) was implemented. A difficulty for the proposed experiment is to keep the high precision of atomic clocks in a Mach-Zehnder interferometer scheme, but in principle this should be possible.


The scheme of Fig.~1 involves a superposition of different states of gravity and, as in previous proposals \cite{Bose:2017nin,Marletto:2017kzi,Saldanha:2026zar,Bose:2026tuv,Bose:2024nhv,Marletto:2024ltk}, its realization would demonstrate a nonclassical behavior of gravity. As shown in Fig. 2, the proper time in a region of space can be outside what is permitted by classical general relativity. 
The scheme of Fig. 3, on the other hand, would demonstrate a superposition of proper times without a superposition of different states of gravity. The fact that quantum systems evolve with respect to proper time, making their evolution sensitive to gravitational time dilation, is well established. Indeed, the effect of gravitational time dilation on atomic clocks has been experimentally resolved even over millimetre-scale height differences \cite{Bothwell2022}. The quantum superposition of distinct proper times, however, has not yet been observed. Such an effect would reveal a new property of the flow of time beyond what can be accounted for by its observer dependence in classical general relativity.

Previous theoretical works predict a reduction in the visibility of the interference pattern in a Mach-Zender interferometer placed in a gravitational field, as represented in the scheme of Fig. 3, due to the different proper times along the two interferometer arms \cite{Zych:2011hu,Pikovski:2013qwa}. Experimentally, however, it could be difficult to establish that a reduction of visibility is indeed caused by this relativistic effect, since there could be many different mechanisms leading to a loss of interference visibility.  The observation of an anomalous proper time, on the other hand, would provide a much more direct verification of a quantum-relativistic effect. Moreover, it is conceivable that the scheme of Fig. 3 can be implemented with current technology, unlike the proposals aimed at demonstrating a quantum superposition of different states of gravity \cite{Bose:2017nin,Marletto:2017kzi,Saldanha:2026zar,Bose:2026tuv,Bose:2024nhv,Marletto:2024ltk}. In this sense, our work presents a promising avenue towards an experimental observation of a genuine general relativistic effect in quantum mechanics. Such an experiment could bring us closer to understanding the possible connections between these two fundamental theories of Physics.

Both authors contributed equally to this work. The authors acknowledge Raul Celistrino Teixeira and Philippe Courteille for very useful discussions. PLS was supported by the Brazilian agency CNPq (Conselho Nacional de Desenvolvimento Científico e Tecnológico). NY benefited from the activities of the COST Actions CA23115: Relativistic quantum information and CA23130: Bridging high and low energies in search of quantum gravity (BridgeQG), funded by COST (European Cooperation in Science and Technology).


\bibliography{qr-time-reversal}

\end{document}